\documentclass[journal]{IEEEtran}
\usepackage{graphicx}

\begin{document}

\title{Level Crossing Rate and Average Fade Duration \\of EGC Systems
with Cochannel Interference \\in Rayleigh Fading}

\author{Zoran~Hadzi-Velkov ~\IEEEmembership{}%
\thanks{}
\thanks{Accepted for IEEE TCOM. The author is with the Faculty of Electrical Engineering,
Ss. Cyril and Methodius University, Karpos 2 bb, 1000 Skopje,
Macedonia (e-mail: zoranhv@feit.ukim.edu.mk)}}%

\markboth{}{Shell \MakeLowercase{\textit{et al.}}: Bare Demo of
IEEEtran.cls for Journals} \maketitle

\begin{abstract}
Both the first-order signal statistics (e.g. the outage
probability) and the second-order signal statistics (e.g. the
average level crossing rate, LCR, and the average fade duration,
AFD) are important design criteria and performance measures for
the wireless communication systems, including the equal gain
combining (EGC) systems in presence of the cochannel interference
(CCI). Although the analytical expressions for the outage
probability of the coherent EGC systems exposed to CCI and various
fading channels are already known, the respective ones for the
average LCR and the AFD are not available in the literature. This
paper presents such analytical expressions for the Rayleigh fading
channel, which are obtained by utilizing a novel analytical
approach that does not require the explicit expression for the
joint PDF of the instantaneous output signal-to-interference ratio
(SIR) and its time derivative. Applying the characteristic
function method and the Beaulieu series, we determined the average
LCR and the AFD at the output of an interference-limited EGC
system with an arbitrary diversity order and an arbitrary number
of cochannel interferers in forms of an infinite integral and an
infinite series. For the dual diversity case, the respective
expressions are derived in closed forms in terms of the gamma and
the beta functions.
\end{abstract}

\begin{keywords}
Level crossing rate, average fade duration, cochannel
interference, equal gain combining, Beaulieu series, Rayleigh
fading.
\end{keywords}


\section{Introduction}

\PARstart{E}{qual} gain combining (EGC) is an important diversity
technique that is often used to mitigate fading in various
wireless communications systems [1]. The EGC has several practical
advantages over other diversity techniques, because it has close
to optimal performance and yet is simple to implement. The outage
probability (OP) is the primary performances measure for all
diversity systems, particularly those exposed to cochannel
interference (CCI) such as the cellular mobile systems. When the
CCI is the predominant noise source at the receiver, its OP
represents the first-order statistical property of the output
signal-to-interference ratio (SIR). The OP of the
interference-limited EGC systems was studied in [2]-[3] and
references therein. Apart from the OP, some aspects in the design
and the analysis of wireless communication systems must also
consider the signal correlation properties, thus necessitating the
determination of its second-order statistical properties: the
average level crossing rate (LCR) and the average fade duration
(AFD). They are used for proper selection of adaptive symbol
rates, interleaver depth, packet length and time slot duration in
various wireless communication systems. While these statistics
have already been determined for the signal envelope at the output
of EGC systems exposed to various fading channels and thermal
noise [4]-[6], the SIR statistics of the EGC systems subject to
CCI have not yet been derived analytically to the best of author's
knowledge. The average LCR and the AFD of the SIR at the output of
selection combining (SC) and maximal-ratio combining (MRC) systems
exposed to CCI and various fading (Rayleigh, Rice and Nakagami)
channels have been reported only recently in [7]-[8], but these
works do not consider the EGC systems. This paper focuses
specifically on an interference-limited coherent EGC system with
an arbitrary diversity order and an arbitrary number of cochannel
interferers, and derives analytical solutions for the average LCR
and the AFD of the output SIR in Rayleigh fading channels.

Section II presents the coherent EGC system model and the channel
model, including the two feasible scenarios for interference
combining. Section III presents the analysis that yields to the
analytical solutions of the OP, the average LCR and the AFD in
forms of an infinite integral and an infinite series. The Section
IV compares the computational burden between these two solutions
and provides several numerical examples that illustrate the
behaviors of the first-order and the second-order signal
statistics. Section V summarizes the main results and concludes
the paper.


\section{System and Channel Models}
We consider a coherent EGC communication receiver with $M$
diversity branches. It is exposed to the transmissions of a single
desired and $N$ interference users, whose signal replicas in each
diversity branch are received over independent identically
distributed (IID) Rayleigh flat fading channels.

In each diversity branch $k$ $\:(1 \leq k \leq M)$, the desired
signal is assumed to have an average power $\Omega_{\rm S}$, while
all interference signals have an equal average power $\Omega_{\rm
I}$. Thus, the channel gains in each branch can be represented as
equivalent complex zero-mean Gaussian random variables (RVs)
$W_{i,k}$; more particulary, $W_{0,k}=X_{0,k} \:
e^{j\theta_{0,k}}$ with variance $\Omega_{\rm S}$ represents the
desired signal in branch $k$, while $W_{i,k}=Y_{i,k} \:
e^{j\theta_{i,k}}$ $(1 \leq i \leq N)$ with variance $\Omega_{\rm
I}$ represents $i$-th interference signal in branch $k$. The
phases of the desired signals $\theta_{0,k}$ and the interference
signal $\theta_{i,k}$ follow the uniform probability distribution
function (PDF) over $[0,2\pi)$, while the respective envelopes
$X_{0,k}=\sqrt{|{W_{0,k}|^2}}$ and $Y_{i,k}=\sqrt{|{W_{i,k}|^2}}$
follow the Rayleigh PDF.

Due to the transmitter/receiver mobility and their relative
velocity, the fading channel introduces time correlation of the
real and imaginary parts of $W_{0,k}$ (i.e., in-phase and
quadrature components of the desired signal) with maximum Doppler
frequency shift $f_{m0}$ in their power spectra. Additionally, the
real and imaginary parts of the channel gains $W_{i,k}$ of each
interfering signal $i$ are also assumed be time correlated with an
identical maximum Doppler frequency shift $f_{mi}$.

In EGC systems, the desired signal replicas in each of the $M$
branches are co-phased, equally weighted, and then coherently
added to give the resultant desired output signal. For the
interference combining, there are two possible scenarios: the
signal replicas originating from any interferer can combine either
incoherently [2, Section III] or coherently [3].

\subsection{Incoherent Interference Combining}
If the interference signals are combined incoherently at the EGC
output, the instantaneous SIR $Z_1$ is determined as [2, Eq. (9)],
\begin{equation}\label{rav1}
Z_1=\frac {(\sum_{k=1}^M X_{0,k})^2}{\sum_{i=1}^N \sum_{k=1}^M
Y_{i,k}^2} \,,
\end{equation}
where the powers (i.e. the squared envelopes) of all interference
signals in all diversity branches are added together. Thus, any
single element in the denominator of (1), $Y_{i,k}^2$, is a
chi-squared RV with 2 degrees of freedom, so the entire
denominator, given by
\begin{equation}\label{rav2}
Y_1^2=\sum_{i=1}^N \sum_{k=1}^M Y_{i,k}^2 \,,
\end{equation}
follows the chi-squared PDF with $2MN$ degrees of freedom,
\begin{equation}\label{rav3}
f_{Y_1^2}(y)= \frac{1}{(\Omega_{\rm I})^{MN}}
\frac{y^{MN-1}}{\Gamma(MN)} \exp\left( -\frac{y}{\Omega_{\rm I}}
\right) \,,
\end{equation}
where $\Gamma(\cdot)$ is the gamma function, defined by
$\Gamma(a)=\int_0^\infty {t^{a-1}e^{-t}dt}$ [14].

\subsection{Coherent Interference Combining}
If the interference signals are combined coherently at the EGC
output, the instantaneous SIR $Z_2$ is determined as [3, Eq. (2)],
\begin{equation}\label{rav4}
Z_2=\frac {(\sum_{k=1}^M X_{0,k})^2}{\sum_{i=1}^N | \sum_{k=1}^M
W_{i,k} |^2} \,,
\end{equation}
where it is assumed that the roll-off factor of the equivalent
baseband communication system is zero. In (4), the complex
interference signals from all branches are first added together
and then squared. Thus, $\sum_{i=1}^M W_{i,k}$ is a complex
Gaussian RV with zero mean and variance $M\Omega_{\rm I}$, while
its squared envelope (i.e. its power) is a chi-squared RV with 2
degrees of freedom. Thus, the denominator in (4), given by
\begin{equation}\label{rav5}
Y_2^2=\sum_{i=1}^N \left | \sum_{k=1}^M W_{i,k} \right | ^2 \,,
\end{equation}
follows the chi-squared PDF with $2N$ degrees of freedom,
\begin{equation}\label{rav6}
f_{Y_2^2}(y)= \frac{1}{(M \Omega_{\rm I})^{N}}
\frac{y^{N-1}}{\Gamma(N)} \exp\left( -\frac{y}{M \Omega_{\rm I}}
\right) \,.
\end{equation}
The PDF of the numerators in (1) and (4), which are the square of
the desired output signal envelope
\begin{equation}\label{rav7}
X=\sum_{k=1}^M X_{0,k} \;,
\end{equation}
is not known in closed form, except for $M=2$. Thus, we revert to
using the {\it characteristic function} (CF) method to arrive at
the desired results.


\section{Average LCR and AFD}

\subsection{Definitions}
We first concentrate on the RV defined as the ratio of the
envelopes of the desired signal $X$ and the equivalent
interference signal $Y_1$ - for incoherent interference combining,
and $Y_2$ - for coherent interference combining,
\begin{equation}\label{rav8}
G_1=\sqrt{Z_1}=\frac{X}{Y_1} \,,
\end{equation}
and
\begin{equation}\label{rav9}
G_2=\sqrt{Z_2}=\frac{X}{Y_2} \,,
\end{equation}
and denoted as the instantaneous {\it envelopes ratio}. We will
first establish the average LCR of the envelopes ratio $G$ and
then readily obtain the average LCR and AFD of the SIR $Z$ based
on (8)-(9). The average LCR of the envelopes ratio $G$ at
threshold $g$ is defined as the rate at which the fading process
crosses level $g$ in the negative direction [1]. It is
mathematically defined by the Rice's formula [1, Eq. (2.106)]
\begin{equation}\label{rav10}
N_G(g)=\int_0^\infty {\dot g}f_{G\dot G}(g,\,\dot g)d\dot g,
\end{equation}
where $\dot G$ denotes the time derivative of $G$, and $f_{G\dot
G}(g,\dot g)$ is the joint PDF of $G$ and $\dot G$. The AFD is
defined as the average time that the envelopes ratio $G$ remains
below the level $g$ after crossing that level in the downward
direction, and is defined by
\begin{equation}\label{rav11}
T_G(g)=\frac{F_G(g)}{N_G(g)} \,,
\end{equation}
where $F_G(\cdot)$ denotes the cumulative distribution function
(CDF) of $G$. Considering (8)-(9), we introduce $g=\sqrt{z}$ into
(10) and (11), and determine the average LCR and the AFD for the
SIR $Z$ at threshold $z$ as $N_Z(z)=N_G(\sqrt z)$ and
$T_Z(z)=T_G(\sqrt z)$, respectively.

\subsection{Characteristic Functions}
The desired signal at the EGC output consists of $M$ Rayleigh RVs
$X_{0,k}$, each having as average power $\Omega=\Omega_{\rm S}$.
Although the PDF of $X$ is not known in closed form (except for $M
= 2$), it is still possible to determine its CF in terms of the
CFs of the constituent $X_{0,k}$s.

For this purpose, we define some general (Nakagami-like) RV $U$
with the PDF given by ($u > 0$),
\begin{equation}\label{rav12}
f(u)= \left( \frac{1}{\Omega} \right)^\alpha
\frac{2u^{2\alpha-1}}{\Gamma(\alpha)} \exp\left(
-\frac{u^2}{\Omega} \right) \,,
\end{equation}
whose CF is given by [2]
\begin{eqnarray}\label{rav13}
\Phi(\omega,\Omega,\alpha) \stackrel{\rm def}{=} \int_0^\infty
{f(u) \exp{(j\omega u)}du} ={}_1F_1
\left(\alpha;\frac12;-\frac{\omega^2}{4}\Omega \right)\nonumber
\end{eqnarray} \vspace{-4.5mm}
\begin{eqnarray}
\qquad \qquad + j\omega\sqrt{\Omega}
\frac{\Gamma(\alpha+1/2)}{\Gamma(\alpha)} {}_1F_1
\left(\alpha+\frac12;\frac32;-\frac{\omega^2}{4}\Omega \right)\,,
\end{eqnarray}
where ${}_1F_1(\cdot;\cdot;\cdot)$ is the confluent (Kummer)
hypergeometric function.

The PDFs of $X_{0,k}$, $Y_1$ and $Y_2$ are given by (12), when
setting $\Omega = \Omega_{\rm S}$, $\alpha=1$ for the $X_{0,k}$;
$\Omega = \Omega_{\rm I}$, $\alpha = MN$ for $Y_1$; and
$\Omega=M\Omega_{\rm I}$, $\alpha=N$ for the $Y_2$, respectively.
Thus, from (13), CFs of $X$, $Y_1$ and $Y_2$ are given by
$\Phi_X(\omega)=[\Phi(\omega,\Omega_{\rm S},1)]^M$,
$\Phi_{Y_1}(\omega)=\Phi(\omega,\Omega_{\rm I},MN)$ and
$\Phi_{Y_2}(\omega)=\Phi(\omega,M\Omega_{\rm I},N)$, respectively.

\subsection{Outage Probability}
The CDF of the envelopes ratio $G$ is determined as [9, Eq. (2)],
\begin{equation}\label{rav14}
F_G(g)=\int_0^\infty {F_X(gy)f_Y(y)dy} \,,
\end{equation}
where $F_X(\cdot)$ is the CDF of the desired signal envelope $X$.
This CDF is expressible in terms of its CF by applying the
Gil-Palaez theorem [10],
\begin{equation}\label{rav15}
F_X(gy)=\frac12 - \frac{1}{2\pi} \int_{-\infty}^\infty
{\frac{\Phi_X(\omega) \exp(-j\omega gy)}{j\omega} d\omega} \,.
\end{equation}
After introducing (15) into (14) and changing the orders of
integration, we have
\begin{eqnarray}
F_G(g)=\frac12 -\frac{1}{2\pi} \int_{-\infty}^\infty
\frac{\Phi_X(\omega)d\omega}{j\omega} \int_0^\infty {f_Y(y)
\exp(-j\omega gy) dy} \nonumber \end{eqnarray} \vspace{-0.25cm}
\begin{eqnarray}
\label{rav16}&=& \frac12 - \frac{1}{2\pi} \int_{-\infty}^\infty
{\frac{\Phi_X(\omega) \Phi_Y^*(g\omega)}{j\omega} d\omega}
\nonumber \\&=& \frac12 - \frac{1}{\pi} \int_{0}^\infty
{\frac{{\rm {Im}} \big \{\Phi_X(\omega) \Phi_Y^*(g\omega) \big
\}}{\omega} d\omega} \,,
\end{eqnarray}
where * denotes conjugate and $\rm{Im} \{\cdot\}$ denotes the
imaginary part of the argument.

The straightforward approach is to estimate (16) by numerical
integration. However, it is also possible to utilize an alternate
approach, which will yield to an infinite series solution of (16).
In [11] Beaulieu derived an infinite series for the PDF and the
CDF of a sum of independent RVs, while [12] gives an alternative
derivation that provided insights into the uses and limitations of
the Beaulieu series. We use this alternative form of the Beaulieu
series [12, Eq. (4b)], and express the CDF of $X$ as
\begin{eqnarray}\label{rav17}
F_X(gx)=\frac12 - \sum_{n=1, n \; \rm{odd}}^\infty {\frac{2 \;
{\rm{Im}}\big \{\Phi_X (n \omega_0) \exp(-j n \omega_0 gx) \big
\}}{n\pi}}\nonumber \end{eqnarray} \vspace{-8.0mm}
\begin{eqnarray}
\qquad \qquad \qquad \qquad \qquad \qquad \qquad \qquad \qquad
\qquad +\Delta_1 \,,
\end{eqnarray}
where $\omega_0=(2\pi/T)$, $T$ is a parameter governing the
sampling rate in the frequency domain and controls the accuracy of
the result, and $\Delta_1$ is an error term that tends to zero for
large $T$. We assume $T$ is large enough to omit this error term.
Introducing (17) over (14) and changing the orders of summation
and integration, we obtain
\begin{equation}\label{rav18}
F_G(g)=\frac12 - \sum_{n=1, n \; \rm{odd}}^\infty {\frac{2 \;
{\rm{Im}} \big \{ \Phi_X (n \omega_0) \Phi_Y^*(n \omega_0 g) \big
\}}{n\pi}} \,.
\end{equation}
The three alternative solutions for the system's outage
probability (i.e. the probability of SIR to fall below a given
threshold $z$) are obtained by setting $g=\sqrt{z}$ into (14),
(16) and (18), which gives
$$\label{rav19a} F_Z(z)=\int_0^\infty {F_X(y
\sqrt{z})f_Y(y)dy} \,, \quad \qquad \qquad \qquad \eqno{(19{\rm
a})}
$$
$$\label{rav19b}F_Z(z)=
\frac12 - \frac{1}{\pi} \int_{0}^\infty { \frac{{\rm{Im}} \big\{
\Phi_X(\omega)  \Phi_Y(-\omega \sqrt{z}) \big \} }{\omega}
d\omega} \,, \eqno{(19{\rm b})}$$ \vspace{-4.5mm}
\begin{eqnarray}\label{rav19c}
F_Z(z)=\frac12 - \frac{2}{\pi} \sum_{n=1}^\infty { \frac{1}{2n-1}}
{\rm{Im}} \big\{ \Phi_X ((2n-1) \omega_0) \quad \nonumber
\end{eqnarray} \vspace{-4.5mm}
$$ \qquad \qquad \qquad \qquad \quad \qquad \times \; \Phi_Y(-(2n-1) \omega_0 \sqrt{z})
\big\} \;,
 \eqno{(19{\rm c})}$$ respectively.

The exact OP can be derived for $M = 2$. The CDF of a sum of the
envelopes of two Rayleigh-faded desired branch signals, $X_{0,1}$
and $X_{0,2}$, is known [13],
\setcounter{equation}{19}
\begin{eqnarray}\label{rav20}
F_X(x)=1 - \exp{\left( -\frac{x^2}{\Omega_{\rm S}} \right)} -
\sqrt{\frac{\pi}{2\Omega_{\rm S}}} \; x  \exp{\left(
-\frac{x^2}{2\Omega_{\rm S}} \right)} \nonumber \\ \times \; {\rm
{erf}} \left( {\frac{x}{\sqrt{2\Omega_{\rm S}}}} \right) \; \,,
\end{eqnarray}
where ${\rm {erf}}(\cdot)$ is the error function. The derivation
of the closed-form solution of (14) for the dual diversity case is
provided in Appendix A, from which the outage probability at
threshold $z$ is determined to be
\begin{eqnarray}\label{rav21}
F_Z(z)=1 - \left( \frac{1}{1+z/\beta} \right)^\alpha -
\frac{\alpha \sqrt{z/(2\beta)}}{[1+z/(2\beta)]^{\alpha+1/2}}
\nonumber \\ \times \; {\rm B}
\left(\frac{1/2}{1+\beta/z};\frac12,\alpha+\frac12 \right) \,,
\end{eqnarray}
where ${\rm B}(\cdot;\cdot,\cdot)$ is the incomplete Beta
function, defined by ${\rm B}_z(a,b) \equiv {\rm
B}(z;a,b)=\int_0^z {t^{a-1} (1-t)^{b-1}dt}$ [14]. In (21),
$(\alpha,\beta)=(2N,\gamma)$ for for incoherent interference
combining, while $(\alpha,\beta)=(N,\gamma/2)$ coherent
interference combining, where $\gamma=\Omega_{\rm S}/\Omega_{\rm
I}$ represents the ratio of the average powers of the desired
signal and a single interference signal in each diversity branch
(also denoted as the average SIR per interferer per branch).

In absence of diversity, $M = 1$, both interference combining
scenarios converge and it is possible to directly solve (19a),
which yields to the classic result for the OP,
\begin{equation}\label{rav22}
F_Z(z)=1 - \frac{1}{(1+z/\gamma)^N} \,.
\end{equation}

\subsection{Average LCR}
In order to determine the average LCR of the random process $G(t)$
by using (8)-(9), one typically needs to establish the joint PDF
of the random processes $G(t)$ and $\dot G(t)$ at any given moment
$t$, $f_{G \dot G}(g,\dot g)$, as according to (10). However, we
utilize an alternative approach, which circumvents explicit
determination of $f_{G \dot G}(g,\dot g)$. From (8)-(9), the time
derivative of the envelopes ratio $G$ is written as
\begin{eqnarray}\label{rav23}
\dot G=\frac{1}{Y}\,\dot X-\frac{X}{Y^2}\,\dot Y=\frac{1}{Y}\,\dot
X-\frac{G}{Y}\,\dot Y\,.
\end{eqnarray}

Conditioned on $Y=y$, the joint PDF $f_{G\dot G}(g,\dot g)$ is
calculated as
\begin{equation}\label{rav24}
f_{G\dot G}(g,\dot g)=\int_0^\infty f_{G\dot G|Y}(g,\dot
g|y)\,f_Y(y)dy\,,
\end{equation}
where $f_Y(y)$ is the PDF of the equivalent interference signal
envelope $Y$. In (24), $f_{G\dot G|Y}(g,\dot g|y)$ is the
conditional joint PDF of $G$ and $\dot G$ given some specified
value of the interference signal envelope $Y=y$, which is
expressed as
\begin{equation}\label{rav25}
f_{G\dot G|Y}(g,\dot g|y)=f_{\dot G|G Y}(\dot g|g,y)\cdot
f_{G|Y}(g|y)\,,
\end{equation}
where $f_{G|Y}(g|y)$ is the conditional PDF of $G$ given $Y=y$.
Because of (8)-(9), it follows $f_{G|Y}(g|y)=y \cdot f_X(gy)$,
where $f_X(\cdot)$ is the PDF of the desired output signal
envelope $X$.

In (25), $f_{\dot G| G Y}(\dot g|g,y)$ is the conditional PDF of
$\dot G$ given some specified values of the envelopes ratio $G=g$
and the interference signal envelope $Y=y$. Considering (23), this
conditional PDF is determined as follows: Conditioned on $G=g$ and
$Y=y$, $\dot G$ is a linear combination of two independent RVs -
the RV representing the time derivative of the desired signal
envelope $\dot X(t)$ and the RV representing the time derivative
of the equivalent interference signal envelope $\dot Y(t)$.

Under certain mathematical conditions, the envelope of the desired
signal $X_{0,k}$ and its respective time derivative $\dot X_{0,k}$
are independent RVs, and, at any given moment $t$, are
characterized by the Rayleigh PDF and the zero-mean Gaussian PDF,
respectively [1]. We conclude that the envelope of the desired
signal at the EGC output $X(t)$ and its time derivative $\dot
X(t)$ are independent, since deriving (7) we get
\begin{equation}\label{rav26}
\dot X = \sum_{k=1}^M {\dot X_{0,k}} \,.
\end{equation}

Hence, $\dot X(t)$ is a zero-mean Gaussian RV with variance equal
to the sum of the variances of the IID Gaussian RVs $\dot
X_{0,k}(t)$ presumed to have equal powers, thus $\sigma_{\dot X}^2
=M\sigma_{\dot X_{0,k}}^2=(\pi f_{m0})^2 M \Omega_{\rm S}$. This
variance is valid for continuous wave (CW) transmission and
two-dimensional isotropic scattering as according to the Clarke's
model [1].

From (2) and (5), it is obvious that the instantaneous
interference powers $Y_1^2$ and $Y_2^2$ can equivalently be
represented as the sum of $MN$ and $N$ IID squared Rayleigh RVs
$R_j$ with average powers $\Omega_{\rm I}$ and $M\Omega_{\rm I}$,
respectively, as $Y^2=\sum_{j=1}^{MN(N)} {R_j^2}$. Finding the
time derivative of both sides of the latter expression and
specifying the values of the constituent Rayleigh RVs (thus fixing
the values of $Y_1$ and $Y_2$), one can easily conclude that both
$\dot Y_1(t)$ and $\dot Y_2(t)$ are zero-mean Gaussian RVs with
variances equal to $\sigma_{\dot Y_1}^2$ and $\sigma_{\dot
Y_2}^2$, respectively, independent of $Y_1(t)$ and $Y_2(t)$ [15,
Section 3.2.1]. The latter conclusion is valid only if the
variances of the time derivative of all constituent Rayleigh RVs
are equal. In this case, assuming two-dimensional isotropic
scattering, $\sigma_{\dot Y_1}^2=(\pi f_{mi})^2 \Omega_{\rm I}$
and $\sigma_{\dot Y_2}^2=(\pi f_{mi})^2 M \Omega_{\rm I}$.

Consequently, $\dot G$ is a zero-mean Gaussian RV with variance
\begin{equation}\label{rav27}
\sigma_{\dot G|GY}^2 = \frac{1}{y^2} \sigma_{\dot X}^2 +
\frac{g^2}{y^2} \sigma_{\dot Y}^2  \,.
\end{equation}

Introducing (24) and (25) into (10), and changing the orders of
integration, we obtain
\begin{eqnarray}\label{rav28}
N_G(g)=\int_0^\infty\dot gd\dot g\int_0^\infty
f_{\dot G|G Y}(\dot g|g,y)\,f_{G|Y}(g|y)\,f_Y(y)dy \nonumber \\
=\int_0^\infty f_{G|Y}(g|y)\,f_Y(y)dy\int_0^\infty\dot g\,f_{\dot
G|G Y}(\dot g|g,y)d\dot g\,\,.
\end{eqnarray}

The inner integral in (28) is calculated by using (27), i.e.,
\begin{equation}\label{rav29}
\int_0^\infty\dot g\,f_{\dot G|G Y}(\dot g|g,y)\,d\dot
g=\frac{\sigma_{\dot G|G Y}}{\sqrt
{2\pi}}=\frac{1}{y}\,\sqrt{\frac{\sigma^2_{\dot
X}+g^2\sigma^2_{\dot Y}}{2\pi}}\,.
\end{equation}

Substituting (29) into (28) and considering $f_{G|Y}(g|y)=y \cdot
f_X(gy)$, we arrive at the important result for the average LCR of
the envelopes ratio $G$ at threshold $g$,
\begin{equation}\label{rav30}
N_G(g)=\sqrt{\frac{\sigma^2_{\dot X}+g^2 \sigma^2_{\dot
Y}}{2\pi}}\,\int_0^\infty f_{X}(gy)\,f_Y(y)dy\,.
\end{equation}

The average LCR of $G$ can also be evaluated in terms of the CFs
of $X$ and $Y$. Namely, after applying the Parseval's theorem over
(30), we directly obtain
\begin{eqnarray}\label{rav31}
N_G(g)= \sqrt{\frac {\sigma_{\dot X}^2+g^2\sigma_{\dot
Y}^2}{2\pi}} \; \frac{1}{2\pi} \int_{-\infty}^\infty {\frac1g \;
\Phi_X\left(\frac{\omega}{g}\right) \;
\Phi_Y^*\left(\omega\right) d\omega} \nonumber \\
=\sqrt{\frac {\sigma_{\dot X}^2+g^2\sigma_{\dot Y}^2}{2\pi}} \;
\frac{1}{2\pi} \int_{-\infty}^\infty {\Phi_X(\omega) \;
\Phi_Y^*(g \omega) d\omega} \qquad \, \nonumber \\
=\sqrt{\frac {\sigma_{\dot X}^2+g^2\sigma_{\dot Y}^2}{2\pi}} \;
\frac{1}{\pi} \int_{0}^\infty {{\rm {Re}} \big \{ \Phi_X(\omega)
\; \Phi_Y^*(g \omega) \big \} d\omega} \,,
\end{eqnarray}
where ${\rm {Re}} \{\cdot\}$ denotes the real part of the
argument.

The straightforward approach to obtain the average LCR of $G$ is
the numerical integration of (31). Alternatively, it is also
possible to calculate the average LCR by using the infinite series
solution after applying the Beaulieu series, similarly to the
derivation of the OP. Namely, the PDF of $X$ is expressed as [12,
Eq. (4a)],
\begin{equation}\label{rav32}
f_X(gy)=\frac4T \sum_{n=1, n \; \rm{odd}}^\infty {{\rm{Re}} \big
\{ \Phi_X (n \omega_0) \exp(-j n \omega_0 gy) \big \} } +\Delta_2
\,,
\end{equation}
where $\Delta_2$ is an error term that tends to zero for large
$T$, as assumed. Introducing (32) over (30) and changing the
orders of summation and integration, we obtain
\begin{eqnarray}\label{rav33}
N_G(g)=\sqrt{\frac {\sigma_{\dot X}^2+g^2\sigma_{\dot Y}^2}{2\pi}}
\; \frac4T \sum_{n=1, n \; \rm{odd}}^\infty { { \rm{Re}} \big \{
\Phi_X (n \omega_0) } \;\quad \nonumber \\
\times \; \Phi_Y^*(n \omega_0 g) \big \} \,.
\end{eqnarray}

Thus, the three alternative solutions for the average LCR of the
SIR $Z$ at threshold $z$ are obtained by setting $g=\sqrt{z}$ into
(30), (31) and (33), yielding
$$\label{rav34a}
N_Z(z)=\sqrt{\frac {\sigma_{\dot X}^2+z \sigma_{\dot Y}^2}{2\pi}}
 \int_0^\infty {f_X(y \sqrt{z})f_Y(y)dy} \,, \qquad \,
\eqno{(34{\rm a})}$$
$$\label{rav34b}
N_Z(z)= \sqrt{\frac
{\sigma_{\dot X}^2+z \sigma_{\dot Y}^2}{2\pi}}
 \frac{1}{\pi} \int_{0}^\infty {{\rm{Re}} \big \{ \Phi_X(\omega)
\Phi_Y(-\omega \sqrt{z}) \big \} d\omega} \,, \eqno{(34{\rm b})}$$
\vspace{-4.0mm}
\begin{eqnarray}\label{rav34c}
N_Z(z)=\sqrt{\frac {\sigma_{\dot X}^2+z \sigma_{\dot Y}^2}{2\pi}}
\;\frac4T \sum_{n=1}^\infty {\rm{Re}} {\big \{ \Phi_X ((2n-1)
\omega_0)} \quad \, \nonumber
\end{eqnarray} \vspace{-2.5mm}
$$ \qquad \; \quad \qquad \qquad \qquad \qquad \times \; \Phi_Y(-(2n-1) \omega_0
\sqrt{z}) \big \} \,, \eqno{(34{\rm c})}$$ respectively.

An exact result can be obtained for the average LCR when $M = 2$.
The PDF of the sum of the envelopes of two Rayleigh-faded desired
branch signals, $X_{0,1}$ and $X_{0,2}$, is known [13],
\setcounter{equation}{34}
\begin{eqnarray}\label{rav35}
f_X(x)=\frac{x}{\Omega_{\rm S}} \exp{\left(
-\frac{x^2}{\Omega_{\rm S}} \right)} +
\sqrt{\frac{\pi}{2\Omega_{\rm S}}} \; \frac{x^2}{\Omega_{\rm
S}} \exp{\left( -\frac{x^2}{2\Omega_{\rm S}} \right)} \quad \nonumber \\
\times \, {\rm {erf}} \left( {\frac{x}{\sqrt{2\Omega_{\rm S}}}}
\right) - \sqrt{\frac{\pi}{2\Omega_{\rm S}}} \; \exp{\left(
-\frac{x^2}{2\Omega_{\rm S}} \right)} \; {\rm {erf}} \left(
{\frac{x}{\sqrt{2\Omega_{\rm S}}}} \right) . \nonumber
\end{eqnarray} \vspace{-5.0mm}
\begin{equation} \end{equation}

The derivation of the closed-form solution of (30) for the dual
diversity case is provided in Appendix A, from which the average
LCR at threshold $z$ for two-dimensional isotropic scattering is
written as
\begin{eqnarray}\label{rav36}
N_Z(z)=\sqrt{\frac {\sigma_{\dot X}^2+z \sigma_{\dot Y}^2}{2\pi}}
\; \sqrt{\frac{1}{\Omega_{\rm S}}}
\frac{\Gamma(\alpha+1/2)}{\Gamma(\alpha)} \frac{1}{1+z/(2\beta)}
 \nonumber \\ \times \left \{
\frac{\sqrt{z/\beta}}{(1+z/\beta)^{\alpha-1/2}}+\sqrt{\frac12} \;
\frac{(\alpha-1/2)z/\beta-1}{(1+z/(2\beta))^\alpha } \right.
\nonumber \\ \times \left. {\rm B}
\left(\frac{1/2}{1+\beta/z};\frac12,\alpha \right) \right \}\,.
\end{eqnarray}

In (36), $(\alpha,\beta)=(2N,\gamma)$ for the incoherent
interference combining, and  $(\alpha,\beta)=(N,\gamma/2)$ for the
coherent interference combining.

If the desired and all interference signals are assumed to have
same maximal Doppler frequency shifts, $f_{m0} = f_{mi}$, (36) is
simplified into
\begin{eqnarray}\label{rav37}
N_Z(z)=f_{m0} \sqrt{\pi} \frac{\Gamma(\alpha+1/2)}{\Gamma(\alpha)}
\frac{1}{\sqrt{1+z/(2\beta)}} \qquad \qquad \quad \nonumber \\
\times \left \{
\frac{\sqrt{z/\beta}}{(1+z/\beta)^{\alpha-1/2}}+\sqrt{\frac12} \;
\frac{(\alpha-1/2)z/\beta-1}{(1+z/(2\beta))^\alpha } \right.
\nonumber \\ \times \left. {\rm B}
\left(\frac{1/2}{1+\beta/z};\frac12,\alpha \right) \right \} \;.
\end{eqnarray}

If the diversity is not employed at the receiver, $M = 1$, both
interference combining scenarios converge, and it is possible to
directly solve (34a) yielding to a well-known result for the
average LCR at threshold $z$ in Rayleigh fading when $f_{m0} =
f_{mi}$ [16, Eq. (17)],
\begin{equation}\label{rav38}
N_Z(z)=f_{m0} \sqrt{2\pi} \frac{\Gamma(N+1/2)}{\Gamma(N)}
\frac{\sqrt{z/\gamma}}{(1+z/\gamma)^N} \,.
\end{equation}

The AFD is calculated as $T_Z(z)=F_Z(z)/N_Z(z)$.

\section{Numeric Examples}
The numeric examples can be calculated either by numerical
integration of (19b) and (34b) or by evaluation of the series
(19c) and (34c). In this Section, we compare these two
approaches/solutions in estimating the OP, the average LCR and the
AFD, and then present some illustrative graphs. It is assumed that
the desired and all interference signal have same maximal Doppler
frequency shifts, $f_{m0} = f_{mi}$.

The first approach requires utilization of a suitable numerical
integration technique embedded in the available computing software
packages, such as MATHEMATICA. The adaptive Gauss-Kronrod
quadrature (GKQ) method is particularly efficient for oscillating
integrands, such as those appearing in (19b) and (34b), which
evaluates the integral at non-equally spaced points (abscissas)
over the integration interval [17]-[18]. The GKQ method
recursively subdivides the integration interval, reusing the
abscissas from the previous iteration as part of the new set of
optimal points, until the result converges to the prescribed
accuracy. The number of abscissas (or, equivalently, the total
number of integrand evaluations) and the number of recursive
subdivisions needed to achieve the desired accuracy are not known
ahead of computation, although the calculations of the new
abscissas in each iteration based on [18] introduce a very low
computational load.

The second approach calculates the numeric examples by truncating
the two infinite series solutions (19c) and (34c) to $L$ non-zero
terms. From the specialized solutions for the OP and the average
LCR for $M = 1$ and $M = 2$, (21), (22), (37) and (38), one can
conclude by induction that $z$ and $\beta$ appear together in the
general expression for an arbitrary $M$, forming the ratio
$z/\beta$. Consequently, the OP is calculated using (19c) and (13)
as
\begin{eqnarray}\label{rav39}
F_Z=\frac12 - \frac{2}{\pi} \sum_{n=1}^L {\frac{1}{2n-1} \;
{\rm{Im}} \big \{ [\Phi ((2n-1) \omega_0,1,1)]^M } \quad \nonumber \\
\times \; \Phi(-(2n-1) \omega_0 \sqrt{z/\beta},1,\alpha) \big \}
\,,
\end{eqnarray}
the normalized average LCR is calculated using (34c) and (13) as
\begin{eqnarray}\label{rav40}
\frac{N_Z}{f_{m0}}=\frac{\sqrt{8\pi}}{T} \sqrt{M+z/\beta} \;
\sum_{n=1}^L {{\rm{Re}} \big \{ [\Phi ((2n-1) \omega_0,1,1)]^M }
\nonumber
\end{eqnarray}\vspace{-5.0mm}
\begin{eqnarray}
\qquad \qquad \qquad \qquad \quad \times \; \Phi(-(2n-1) \omega_0
\sqrt{z/\beta},1,\alpha) \big \} ,
\end{eqnarray}

\begin{figure}
\centering
\includegraphics[width=3.5in]{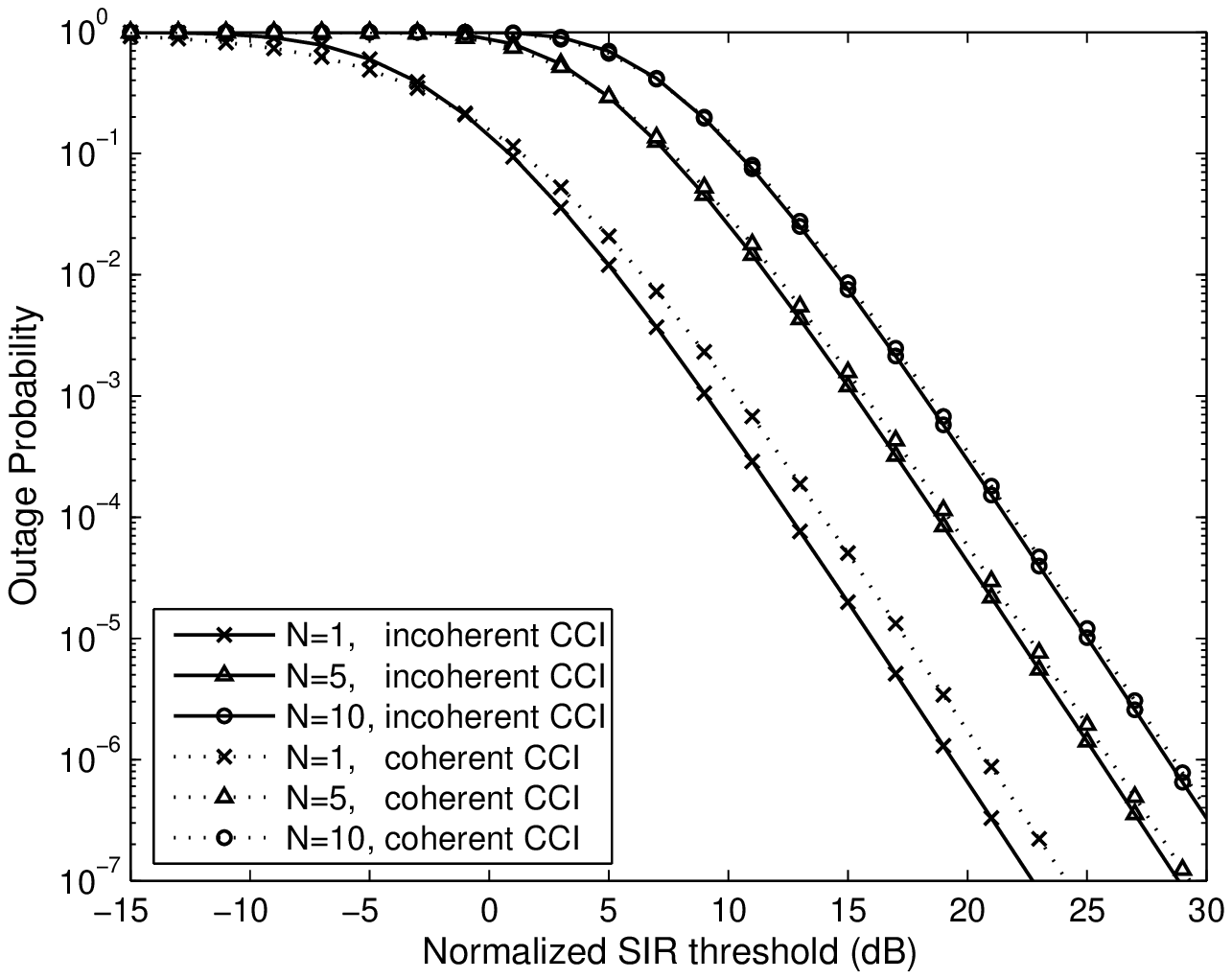}
\begin{center} \footnotesize (a) Behavior of OP \end{center}
\label{fig_1a}
\end{figure}
\begin{figure}
\centering
\includegraphics[width=3.5in]{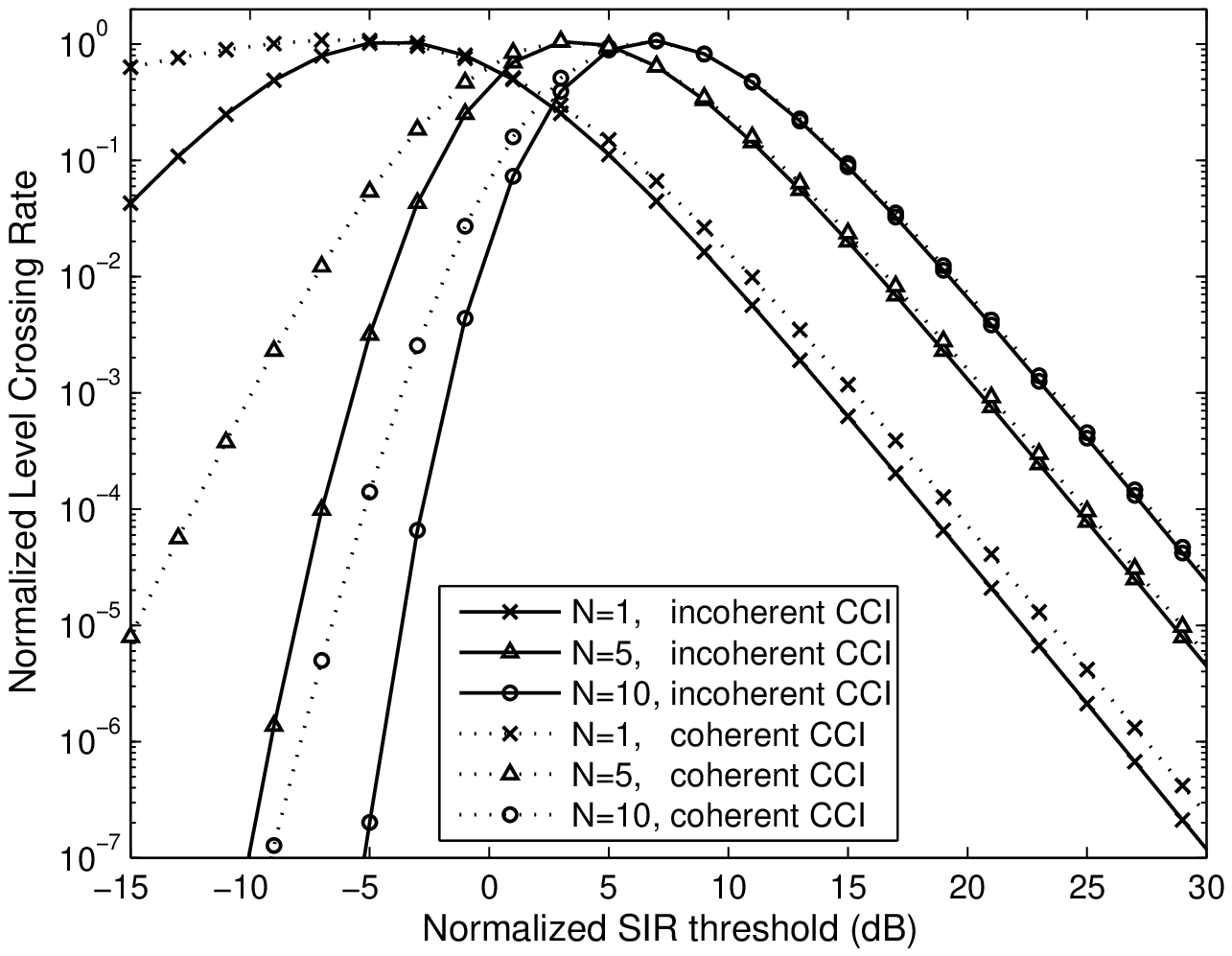}
\begin{center} \footnotesize (b) Behavior of average LCR \end{center}
\label{fig_1b}
\end{figure}
\begin{figure}
\centering
\includegraphics[width=3.5in]{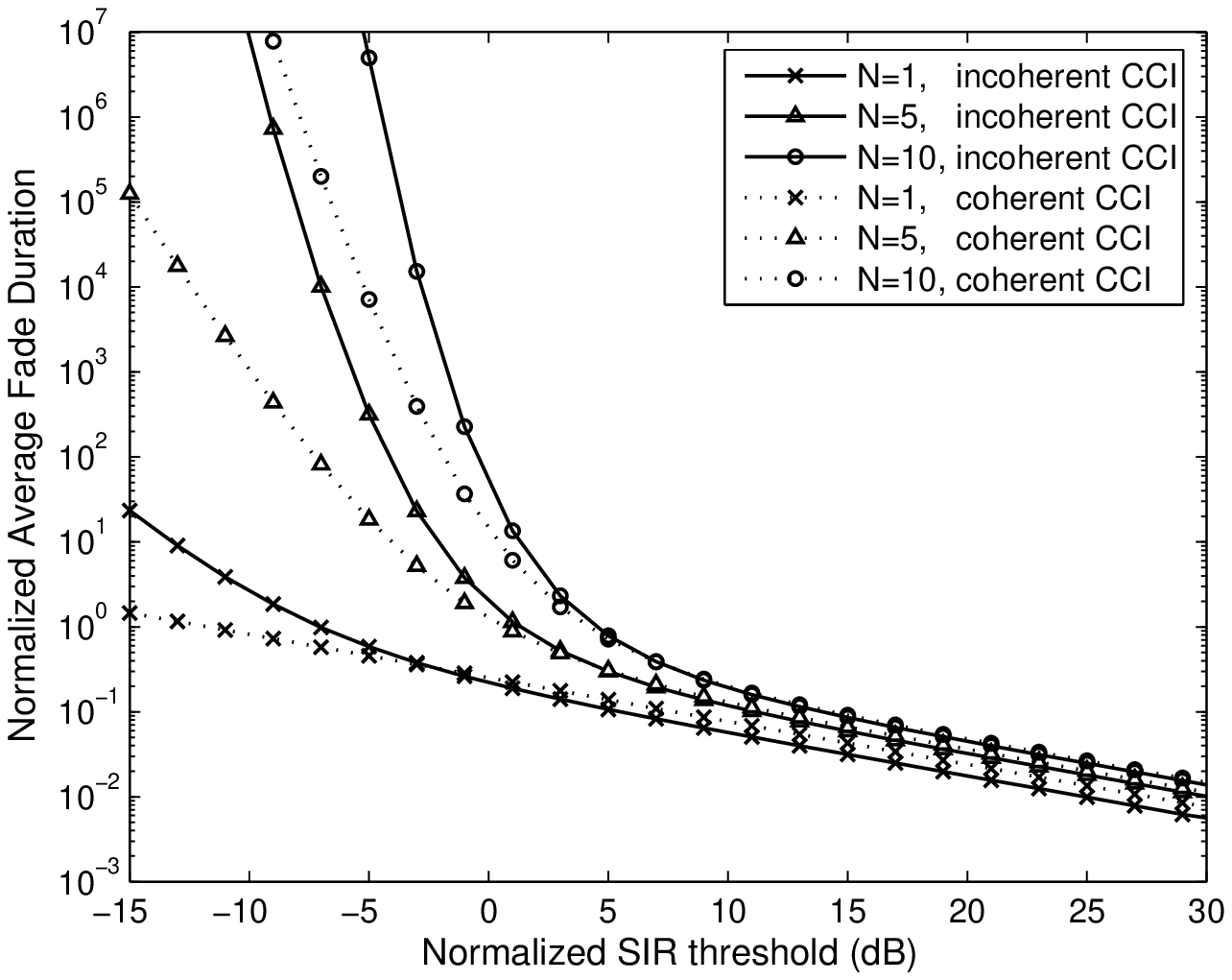}
\begin{center} \footnotesize (c) Behavior of AFD  \end{center}
\vspace{-3mm}
\setcounter{figure}{0} \caption {First-order and second-order EGC
output signal statistics vs. SIR thresholds for various numbers of
interferers when $M = 3$} \label{fig_1c}
\end{figure}
\begin{figure}
\centering
\includegraphics[width=3.5in]{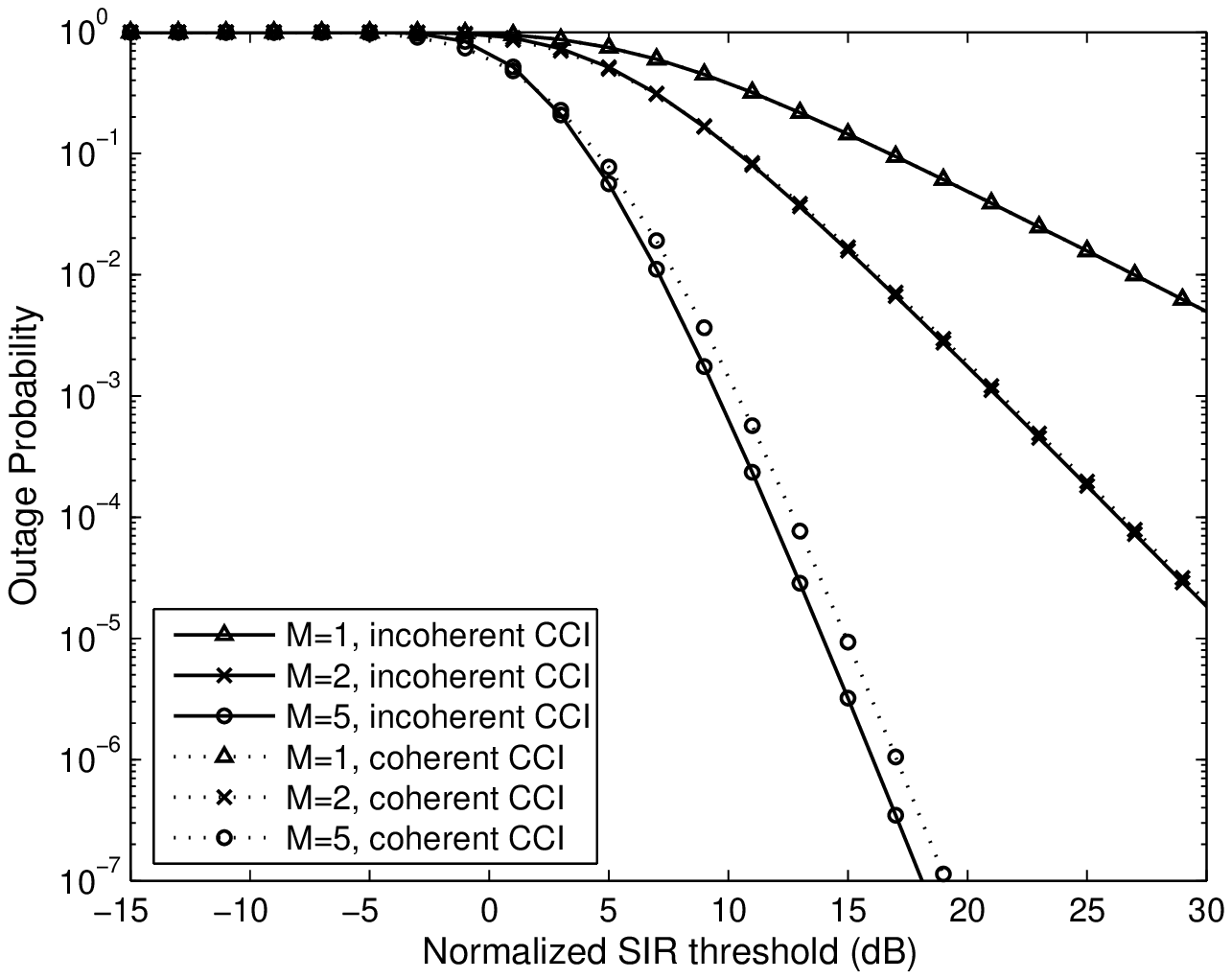}
\begin{center} \footnotesize (a) Behavior of OP \end{center}
\label{fig_2a}
\end{figure}
\begin{figure}
\centering
\includegraphics[width=3.5in]{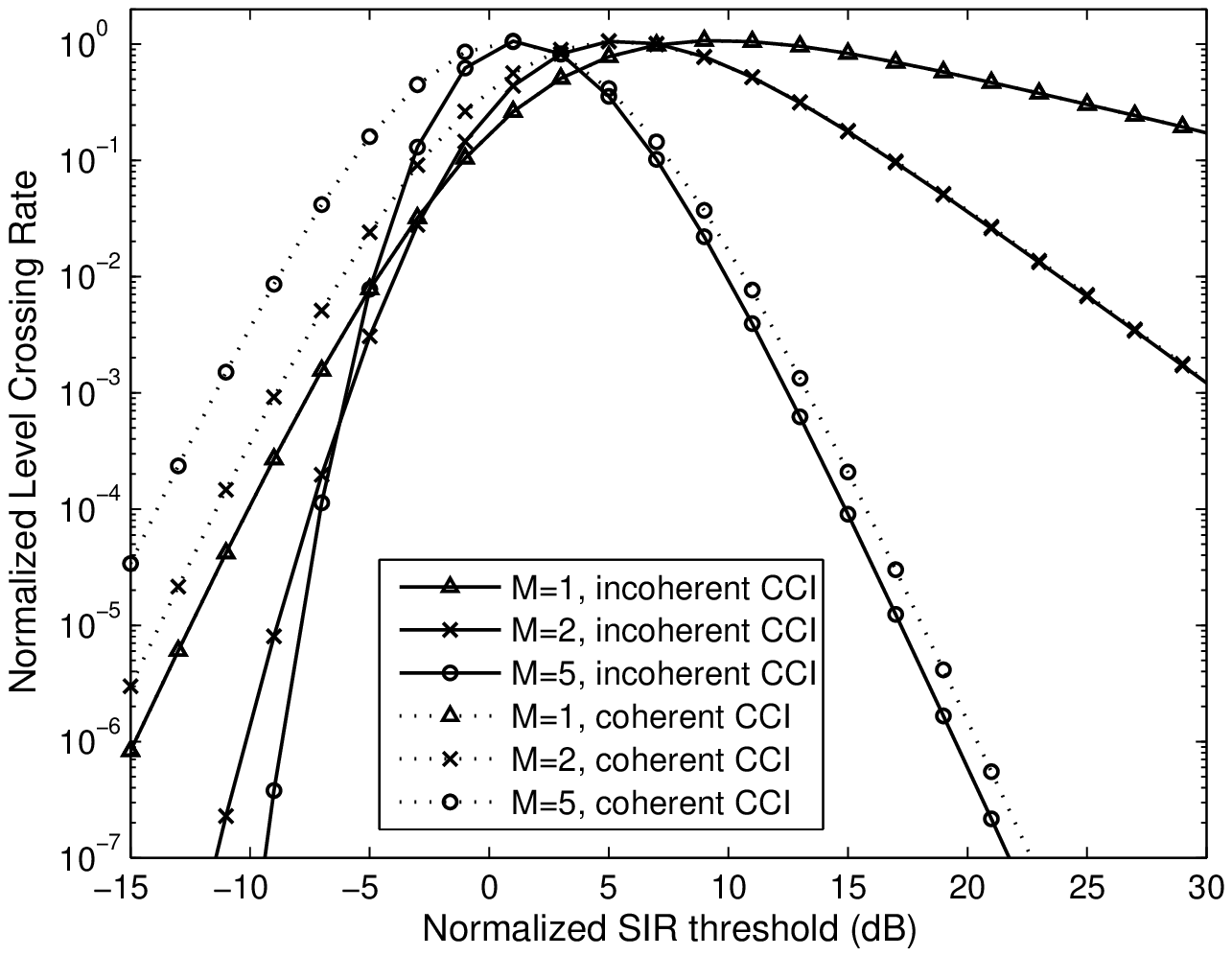}
\begin{center} \footnotesize (b) Behavior of average LCR \end{center}
\label{fig_2b}
\end{figure}
\setcounter{figure}{1}
\begin{figure}
\centering
\includegraphics[width=3.5in]{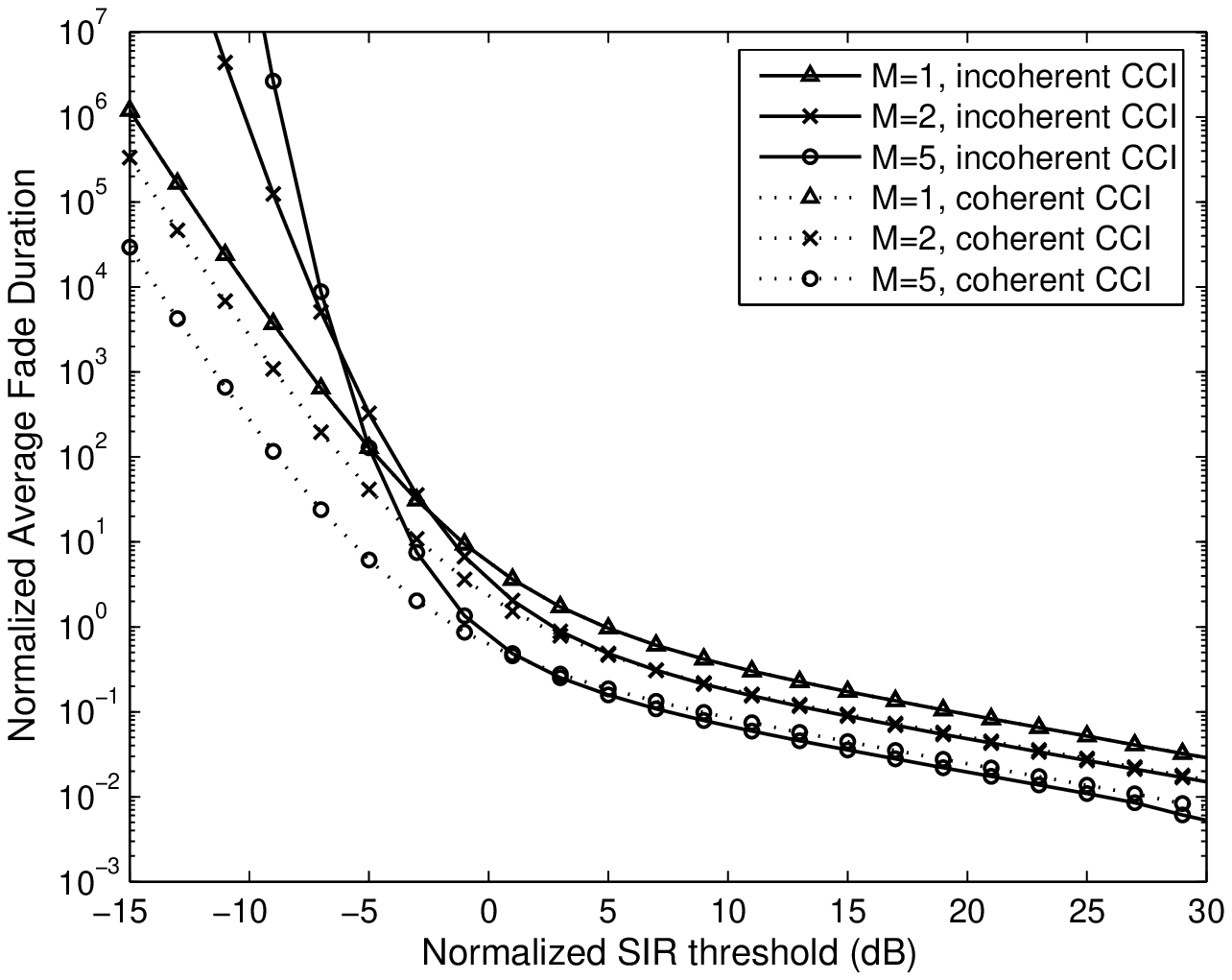}
\begin{center} \footnotesize (c) Behavior of AFD  \end{center}
\vspace{-3mm}
\setcounter{figure}{1} \caption{First-order and second-order EGC
output signal statistics vs. SIR thresholds for various diversity
orders when $N = 5$} \label{fig_2c}
\end{figure}
\begin{figure}
\centering
\includegraphics[width=3.5in]{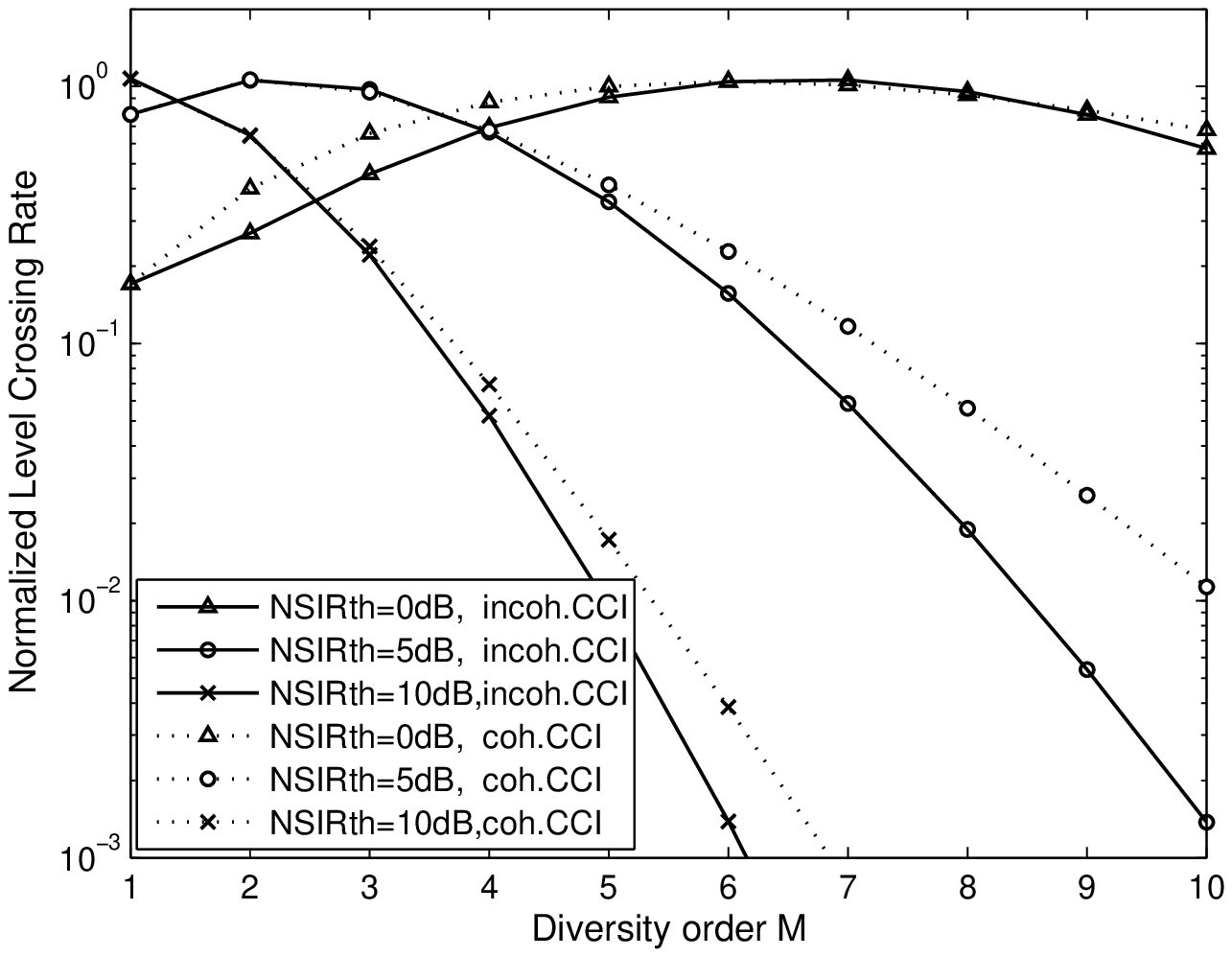}
\begin{center} \footnotesize (a) Behavior of average LCR \end{center}
\label{fig_3a}
\end{figure}
\setcounter{figure}{2}
\begin{figure}
\centering
\includegraphics[width=3.5in]{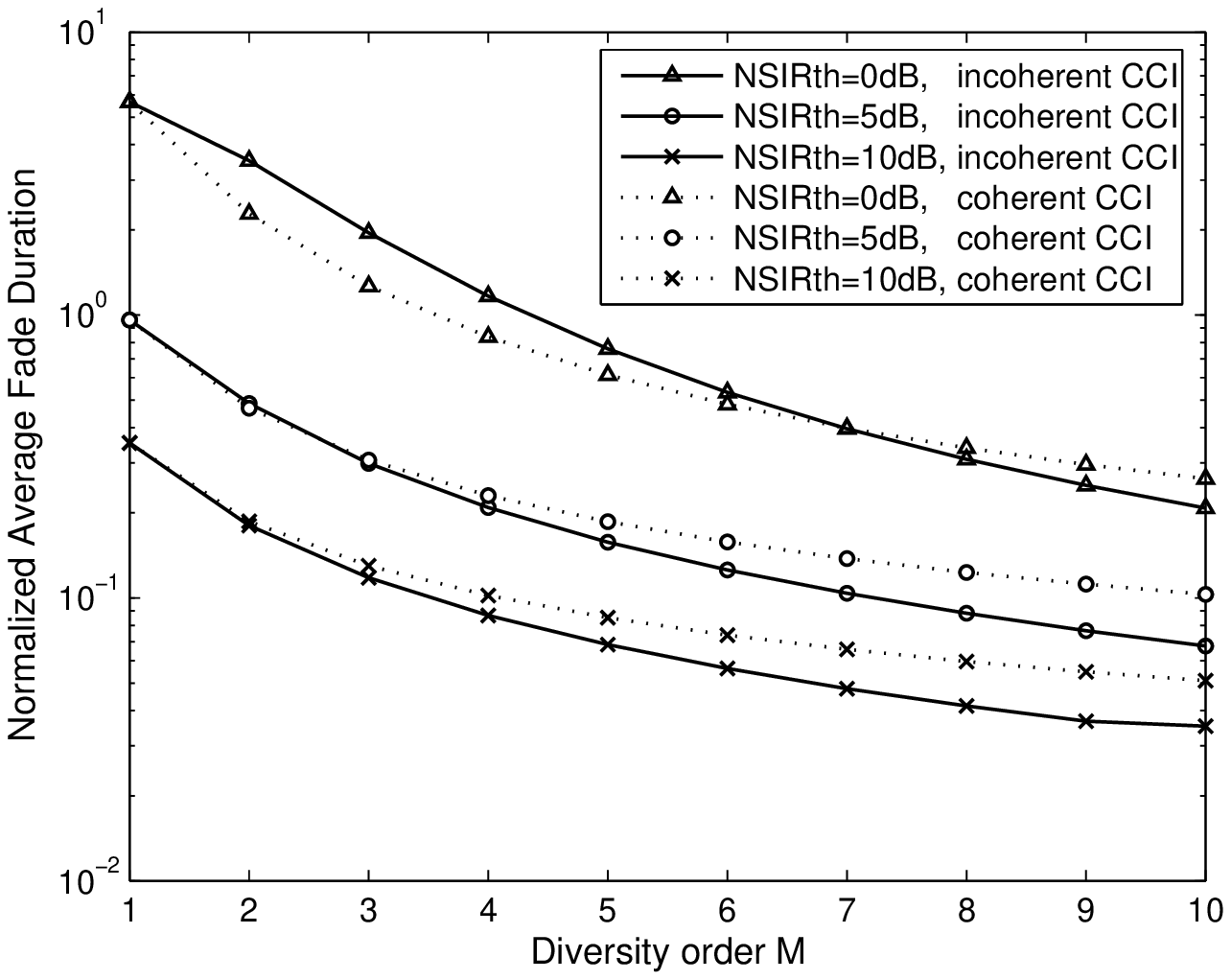}
\begin{center} \footnotesize (b) Behavior of AFD  \end{center}
\vspace{-3mm}
\setcounter{figure}{2} \caption{Second-order EGC output signal
statistics vs. diversity order for various normalized SIR
thresholds when $N = 5$} \label{fig_3b}
\end{figure}
\noindent and the normalized AFD, $f_{m0}F_Z/N_Z$ is calculated
from the ratio of (39) and (40). Depending on the presumed
scenario, $(\alpha,\beta)=(MN,\gamma)$ - for the incoherent
interference combining, and $(\alpha,\beta)=(N,\gamma/M)$ - for
the coherent interference combining, while the normalized SIR
threshold (NSIRth) is determined as $\gamma/z = \Omega_{\rm
S}/(\Omega_{\rm I}z)$. Note that (39) and (40) actually estimate
(19b) and (34b), respectively, by sampling their integrands at
equally spaced abscissas, whereas their number $L$ and locations
(odd multiples of $\omega_0$) are given ahead of computation.
There is a tradeoff between the absolute accuracy and the
selection of $T$ and $L$. A larger value of $T$ results in greater
accuracy, but more nonzero terms $L$ must be used [11].
\begin{figure}
\centering
\includegraphics[width=3.5in]{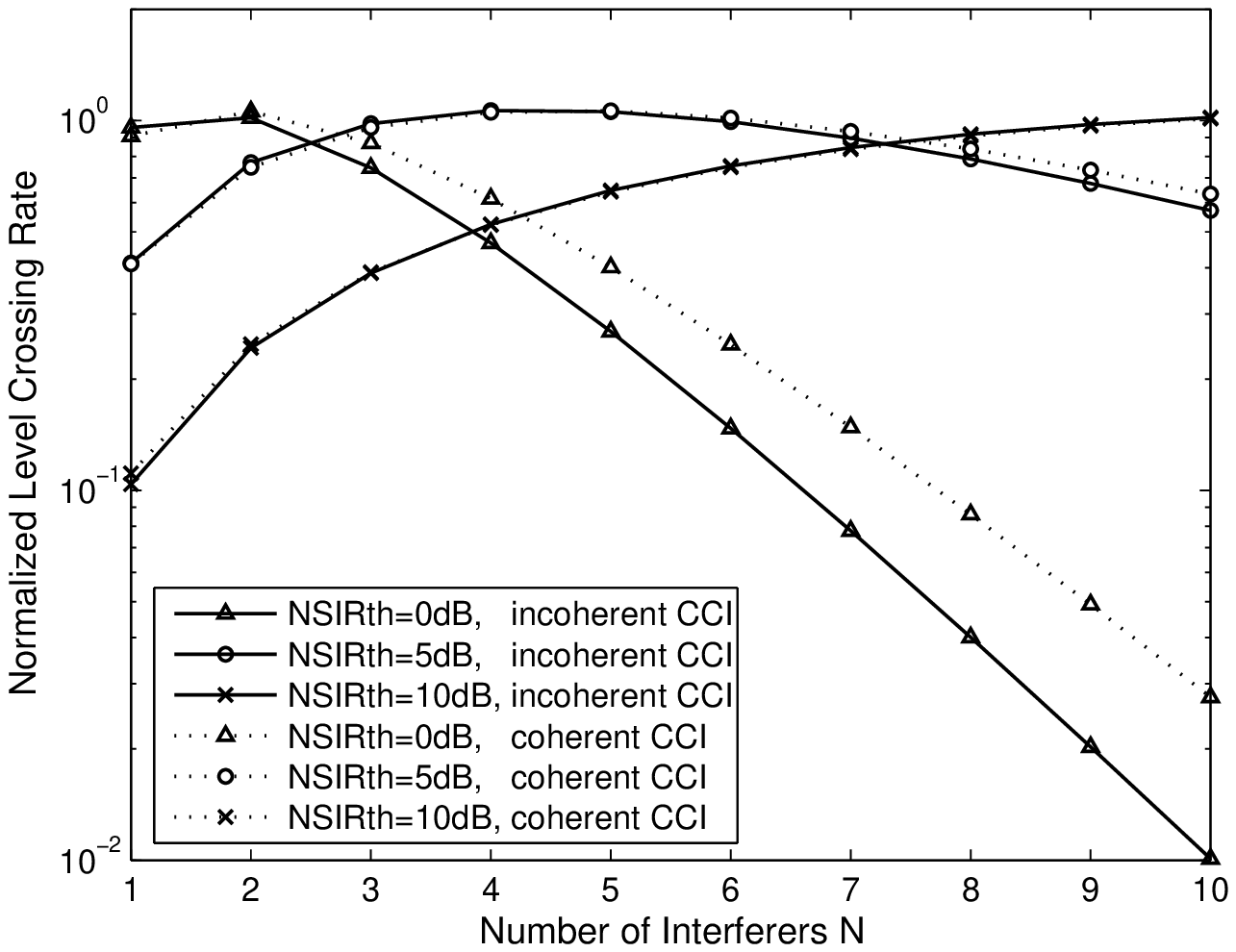}
\begin{center} \footnotesize (a) Behavior of average LCR \end{center}
\label{fig_4a}
\end{figure}
\setcounter{figure}{3}
\begin{figure}
\centering
\includegraphics[width=3.5in]{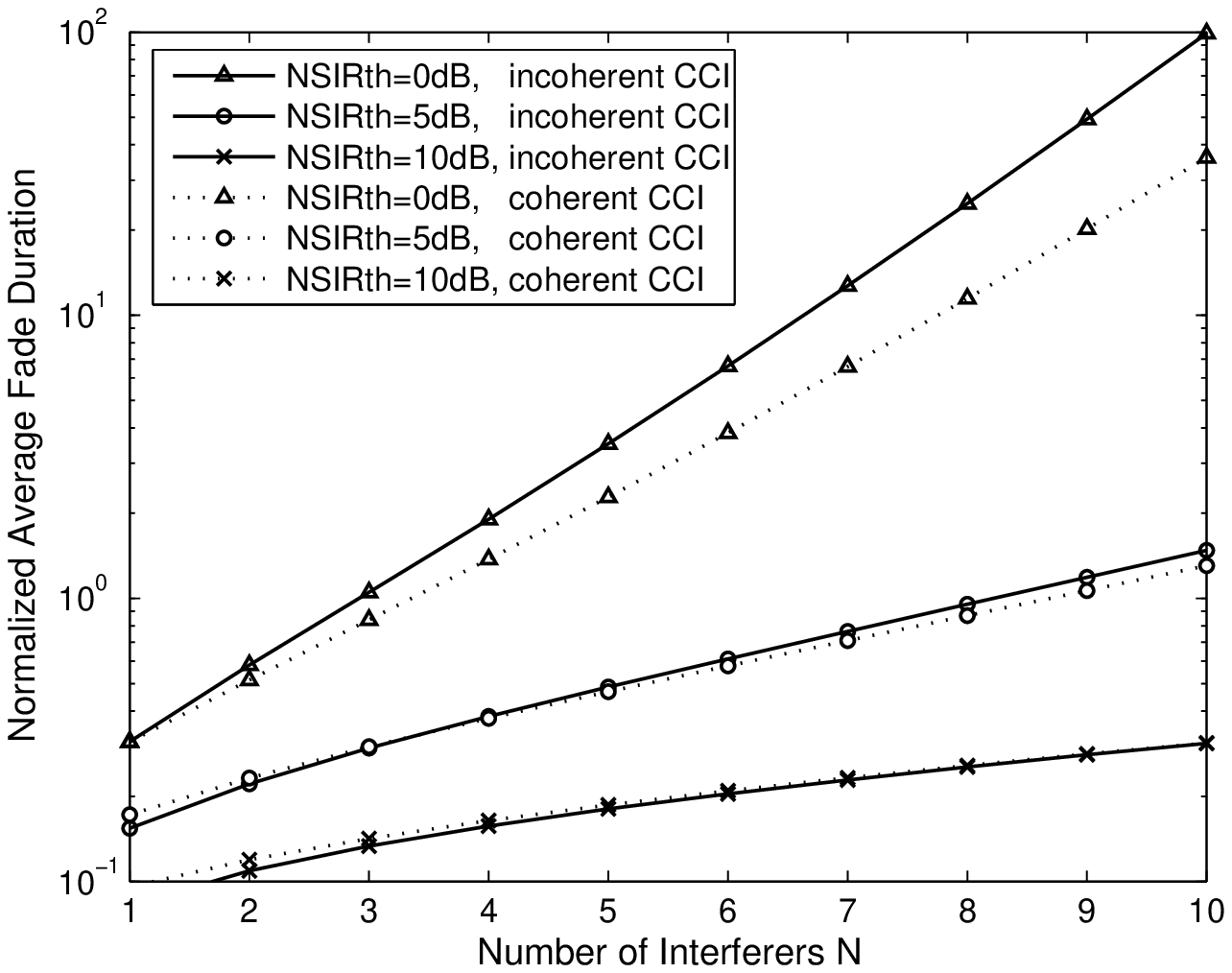}
\begin{center} \footnotesize (b) Behavior of AFD  \end{center}
\vspace{-3mm}
\setcounter{figure}{3} \caption{Second-order EGC output signal
statistics vs. number of interferers for various normalized SIR
thresholds when $M = 2$} \label{fig_4b}
\end{figure}
Using MATHEMATICA, we compared the computational burden between
the two approaches/solutions by calculating the same numerical
examples with same prescribed absolute accuracy of $\pm 10^{-8}$.
In utilizing the first approach, we set the target accuracy into
the computing software, and, for a given set of input parameters
(NSIRth, $M$ and $N$), obtain the integration result and the
respective value of the built-in variable that counts the number
of integrand evaluations. In utilizing the second approach, the
values of $T$ and $L$ needed to achieve the target accuracy are
obtained empirically by trying multiple combinations of $T$ and
$L$, thus yielding to typical values of $T$ between 40 and 100,
and $L$ - between 100 and 200.

For a given selection of $T$ and $L$, (39) has a significantly
better rate of convergence then (40), so the accuracy of (40)
determines the accuracy of both the normalized AFD and normalized
average LCR. Similarly, the GKQ of the OP (19b) produces a
numerical result with better convergence rate and less integrand
evaluations compared to the GKQ of the average LCR (34b).

We established that the first approach introduces higher
computational load, thus requiring longer computation times.
Namely, for the range of the input parameters shown on Figs. 1, 2
and 3, a single numeric integration using the GKQ method requires
between 100 and 500 integrand evaluations to estimate a single
value of the OP or the average LCR. In the same range of the input
parameters, the truncated Beaulieu series typically require fewer
number of integrand evaluations ($L$ nonzero terms) to achieve the
same accuracy, thus yielding to shorter computation times to
obtain the respective results.

It is also observed that the increase of $\alpha$ adds to the
computational burden of the GKQ of both (19b) and (34b), since
their integrands become more rapidly oscillatory (particularly
emphasized for the average LCR calculations), thus requiring more
integrand evaluations (increasing toward 500) to achieve the
desired accuracy. The number of required nonzero terms $L$ in the
Beaulieu series increases (toward 200) for lower NSIRth (i.e.
higher threshold $z$), but rapidly decreases with the increase of
the diversity order $M$. Compared to the first approach, however,
the computational loads of the Beaulieu series in estimating OP
and average LCR are less dependent from their input parameters.

Figs. 1 and 2 depict the OP, the normalized average LCR and the
normalized AFD versus the NSIRth, with $N$ and $M$ appearing as
curve parameters ($M = 3$, $N =$ 1, 5 and 10 in Fig. 1, and $M =$
1, 2 and 5, $N =$ 5 in Fig. 2). Note that if NSIRth $<$ 0 dB then
$z > \Omega_{\rm S}/\Omega_{\rm I}$, while if NSIRth $>$ 0 dB then
$z < \Omega_{\rm S}/\Omega_{\rm I}$.

As expected, Figs. 1a and 2a show that the OP is a monotonically
decreasing function from the NSIRth, and that its values match [2,
Fig. 1] and [3, Figs. 2 and 3] in the respective NSIRth ranges for
given $N$ and $M$. Figs. 1b and 2b show that the average LCR
reaches its maximum for some specific NSIRth, whose value depends
on the selection of $M$ and $N$. Figs. 1c and 2c show that the AFD
decreases by the increase of the NSIRth (i.e. by the decrease of
the threshold $z$).

It is also obvious that, for a given values of NSIRth, $M$ and
$N$, the average LCR and the AFD curves for incoherent and
coherent interference combining almost coincide when the NSIRth is
above the value that maximizes the average LCR, while they differ
below this value.

Figs. 3 and 4 depict the influence of the diversity order $M$ and
the number of interferers $N$ over the average LCR and the AFD for
different values of NSIRth. Depending on whether NSIRth is set
below or above the value that maximizes the average LCR, the
average LCR may increase and/or decrease by increasing $M$ (Fig.
3a), while the AFD monotonically decreases (Fig. 3b). The average
LCR may also increase and/or decrease by increasing $N$ (Fig. 4a),
while the AFD monotonically increases (Fig. 4b).

Extensive Monte Carlo simulations conducted in MATLAB have
validated all numeric examples presented in this Section.

\section{Conclusion}
This paper derived the analytical expressions for the average LCR
and the AFD of coherent EGC wireless communication systems subject
to CCI and Rayleigh fading.

The solutions for the average LCR were derived by a novel
analytical approach that circumvents the necessity of finding the
explicit expression for the joint PDF of the instantaneous SIR and
its time derivative. They have been expressed in forms of an
infinite integral solution and an infinite series solution,
assuming IID equal-powered interference signals' replicas and IID
equal-powered desired signal replicas in each diversity branch.
The infinite series solutions were determined for an arbitrary
diversity order after successively applying the CF method, the
Parseval's theorem and the Beaulieu series over the integral
expressions for the OP and average LCR. Compared to the numerical
integration method commonly implemented in the computing software
packages, we concluded that the Beaulieu series solutions
introduce less computational burden and minor sensitivity to the
input parameters. For the dual diversity case, the average LCR and
the AFD were determined as exact closed-form solutions in terms of
the gamma and the beta functions.

For the interference-limited EGC systems, the desired branch
signals coherently combine, while the branch signals from each
interferer can combine either coherently or incoherently. Our
analytical solutions incorporate both combining scenarios,
yielding to somewhat different numeric values for the average LCR
and the AFD. The differences are more evident when the SIR
threshold is set above the average SIR per interferer per branch.

One can further alleviate the assumption for the equal branch
powers of the desired signal replicas. It is straightforwardly
obvious from (26) that same analytical approach is also applicable
for determination of the average LCR in the case of unequal branch
powers of the desired signal. The derivation of the respective
solutions is trivial and omitted in this work.


\useRomanappendicesfalse
\appendices
\section{}
Introducing the expressions for the CDF of
$X$ (20) and the PDF of $Y$ (12) into (14), and also using [14,
Eq. 3.478(1), 6.286(1)], one obtains
\renewcommand{\theequation}{\thesection.\arabic{equation}}
\setcounter{equation}{0}
\begin{eqnarray}\label{ravA.1}
F_G(g)=1 - \left(\frac{1}{1+g^2/\beta}\right)^\alpha -
\frac{\alpha g^2/\beta}{[1+g^2/(2\beta)]^{\alpha+1}} \nonumber \\
\times \; {}_2F_1 \left(
\frac12;1+\alpha;\frac32;-\frac{1}{1+2\beta/g^2} \right) \,,
\end{eqnarray}
where ${}_2F_1(a;b;c;z)$ is the Gaussian hypergeometric function
[14]. We then successively apply transformations [14]
\begin{equation}\label{ravA.2}
{}_2F_1(a;b;c;z) = \left( \frac{1}{1-z} \right)^b \; {}_2F_1
\left( c-a;b;c;\frac{z}{z-1} \right)
\end{equation}
and
\begin{equation}\label{ravA.3}
{}_2F_1(1;b;c;z) =  \frac{c-1}{z^{c-1}} (1-z)^{c-b-1} \; {\rm B} (
z;c-1,b-c+1 )
\end{equation}
over (A.1) and obtain
\begin{eqnarray}\label{ravA.4}
F_G(g)=1 - \left( \frac{1}{1+g^2/\beta} \right)^\alpha -
\frac{\alpha \sqrt{g^2/(2\beta)}}{[1+g^2/(2\beta)]^{\alpha+1/2}}
\nonumber \\ \times \; {\rm B}
\left(\frac{1/2}{1+\beta/g^2};\frac12,\alpha+\frac12 \right) \,.
\end{eqnarray}
The result (21) is obtained directly from (A.4) by setting
$g=\sqrt{z}$.

\pagebreak{}Introducing the expressions for the PDF of $X$ (35)
and the PDF of $Y$ (12) into (30), and also using [14, Eq.
3.478(1), 6.286(1)], one obtains
\begin{eqnarray}\label{ravA.5}
N_G(g)=\sqrt{\frac {\sigma_{\dot X}^2+g^2 \sigma_{\dot
Y}^2}{2\pi}} \; \left \{ \sqrt{\frac{1}{\Omega_{\rm S}}}
\frac{\Gamma(\alpha+1/2)}{\Gamma(\alpha)} \qquad \qquad \quad
\right. \nonumber
\end{eqnarray} \vspace{-5.0mm}
\begin{eqnarray}
\times \frac{g^2/\sqrt{\beta}}{(1+g^2/\beta)^{\alpha+1/2}}
+\sqrt{\frac{1}{\Omega_{\rm S}}}
\frac{\Gamma(\alpha+3/2)}{\Gamma(\alpha)} \qquad \qquad \qquad
\nonumber
\end{eqnarray}\vspace{-5.0mm}
\begin{eqnarray}
\times \; \frac{(g/\sqrt{\beta})^3}{(1+g^2/(2\beta))^{\alpha+3/2}}
\; {}_2F_1
\left(\frac12;\frac32+\alpha;\frac32;-\frac{1}{1+2\beta/g^2}
\right) \nonumber
\end{eqnarray}\vspace{-5.0mm}
\begin{eqnarray}
- \sqrt{\frac{1}{\Omega_{\rm S}}}
\frac{\Gamma(\alpha+1/2)}{\Gamma(\alpha)}
\frac{g/\sqrt{\beta}}{(1+g^2/(2\beta))^{\alpha+1/2}} \qquad \qquad
\qquad \quad \nonumber
\end{eqnarray}\vspace{-5.0mm}
\begin{eqnarray}
\left. \quad \qquad \qquad \times \; {}_2F_1
\left(\frac12;\frac12+\alpha;\frac32;-\frac{1}{1+2\beta/g^2}
\right) \right \} \,.
\end{eqnarray}

After successively applying both (A.2) and (A.3) over (A.5), one
obtains
\begin{eqnarray}\label{ravA.6}
N_G(g)=\sqrt{\frac {\sigma_{\dot X}^2+g^2 \sigma_{\dot
Y}^2}{2\pi}} \; \left \{ \sqrt{\frac{1}{\Omega_{\rm S}}}
\frac{\Gamma(\alpha+1/2)}{\Gamma(\alpha)} \qquad \qquad \qquad
\right. \nonumber
\end{eqnarray}\vspace{-5.0mm}
\begin{eqnarray}
\times \; \frac{g^2/\sqrt{\beta}}{(1+g^2/\beta)^{\alpha+1/2}}
+\sqrt{\frac{1}{2\Omega_{\rm S}}}
\frac{\Gamma(\alpha+3/2)}{\Gamma(\alpha)} \qquad \qquad \nonumber
\end{eqnarray}\vspace{-5.0mm}
\begin{eqnarray}
\times \; \frac{g^2/\beta}{(1+g^2/(2\beta))^{\alpha+1}} \; {\rm B}
\left(\frac{1/2}{1+\beta/g^2};\frac12,1+\alpha \right) \qquad
 \nonumber
\end{eqnarray}\vspace{-5.0mm}
\begin{eqnarray}
- \sqrt{\frac{1}{2\Omega_{\rm S}}}
\frac{\Gamma(\alpha+1/2)}{\Gamma(\alpha)}
\frac{1}{(1+g^2/(2\beta))^{\alpha}} \qquad \qquad \qquad \nonumber
\end{eqnarray}\vspace{-7.5mm}
\begin{eqnarray}
\left. \qquad \qquad \qquad \qquad \qquad \times \; {\rm B}
\left(\frac{1/2}{1+\beta/g^2};\frac12,\alpha \right) \right \}\,.
\end{eqnarray}
We further apply the following identity [14]
\begin{equation}\label{ravA.7}
{\rm B}(z;a,b+1) = \frac{z^\alpha (1-z)^b}{a+b}+ \frac{b}{a+b} \;
{\rm B} \left(z;a,b \right)
\end{equation}
over (A.6), and obtain
\begin{eqnarray}\label{ravA.8}
N_G(g)=\sqrt{\frac {\sigma_{\dot X}^2+g^2 \sigma_{\dot
Y}^2}{2\pi}} \; \left \{ \sqrt{\frac{1}{\Omega_{\rm S}}}
\frac{\Gamma(\alpha+1/2)}{\Gamma(\alpha)} \qquad \qquad \qquad
\right. \nonumber
\end{eqnarray}\vspace{-5.0mm}
\begin{eqnarray}
\times \frac{1}{1+g^2/(2\beta)}
\frac{g/\sqrt{\beta}}{(1+g^2/\beta)^{\alpha-1/2}}
+\sqrt{\frac{1}{2\Omega_{\rm S}}}
\frac{\Gamma(\alpha+1/2)}{\Gamma(\alpha)} \nonumber
\end{eqnarray}\vspace{-6.0mm}
\begin{eqnarray}
\left. \quad \qquad \times \;
\frac{(\alpha-1/2)g^2/\beta-1}{(1+g^2/(2\beta))^{\alpha+1}} \;
{\rm B} \left(\frac{1/2}{1+\beta/g^2};\frac12,\alpha \right)
\right \}\,.
\end{eqnarray}
The result (36) is obtained directly from (A.8) after setting
$g=\sqrt{z}$.



\section*{Acknowledgement}
The author wishes to thank the Editor and the anonymous reviewers
for their useful comments that improved the quality of this paper.



\end{document}